\documentclass{article}

\usepackage{graphicx} 
\usepackage{subcaption}
\usepackage{color}

\title{A Scale-Invariant Theory of the Universe}
\author{
Julian Barbour\thanks{julian.barbour@physics.ox.ac.uk}\\
\small College Farm, The Town, South Newington,\\
\small Banbury OX15 4JG, United Kingdom
\and
Maria I. R. Lourenço\thanks{fc56407@alunos.fc.ul.pt}\\
\small Instituto de Astrofísica e Ciências do Espaço,\\
\small Faculdade de Ciências da Universidade de Lisboa,\\
\small Edifício C8, Campo Grande,\\
\small P-1749-016 Lisbon, Portugal
}
\date{}

\begin{document}

\maketitle

\begin{abstract}
Modern physics has achieved extraordinary empirical success while retaining much of the absolute, unobservable structure introduced by Newton, largely without questioning its necessity. We investigate how far this structure can be eliminated by adopting a relational ontology guided by Leibniz's principle of sufficient reason. Removing absolute position, orientation, time and, finally, scale leads naturally to a formulation of the gravitational $N$-body problem where only dimensionless ratios are physically meaningful. Within this framework, the scale-invariant variety $V$ becomes a central quantity, providing a measure of structure, a natural ordering of shapes, and an emergent gravitational arrow of time. We argue that the resulting formulation unifies classes of Newtonian solutions previously regarded as distinct, uncovering a possible new symmetry, suggests a notion of explanation based on timeless spatial correlations rather than temporal evolution, and points towards a more economical ontology. Although developed in the context of Newtonian gravity, the principles proposed here may also offer a new perspective on general relativity and quantum mechanics.
\end{abstract}

\section{Introduction}

In this article, we present the key elements of a relational, holistic, and shape-based theory of the universe, review its principal results, and discuss their possible interpretations and its potential implications for general relativity and quantum mechanics. Our aim is to make explicit what a theory of shapes is capable of achieving and to explore the conceptual shift it entails with respect to established physics.

We begin with the concept of symmetry, which lies at the heart of modern physics. Indeed, symmetry is so deeply embedded in the subject that one could describe physics as the science that identifies mathematical patterns in nature. From translational and rotational invariance in classical mechanics to Lorentz invariance and gauge symmetries in modern physics, symmetry principles have shaped our deepest physical theories. Yet, despite its conceptual simplicity, there is one notable exception: \emph{scale invariance}. It is realised, for example, when all lengths are multiplied by the same factor while nothing observable changes. Although scale invariance plays a  role in several areas of modern physics, it has not attained the status of a universal first principle governing the structure of physical law. Poincaré had already emphasised the empirical indistinguishability of a global rescaling, arguing in \cite{P}, that if during a single night all sizes---his own, his bed, his house, the Earth, and so on---were rescaled by the same factor, the world would appear exactly the same to every observer. Surprisingly, this fact has largely escaped notice in the mathematical description of nature. For example, the Newtonian potential is invariant under translations and rotations, but not under dilatations.

Taking scale invariance as a foundational principle invites us to assign greater significance to ratios, or, in other words, to \emph{shapes}. Consider a triangle. Its size may be increased or decreased without altering the ratios between its sides or its internal angles. These quantities are therefore scale-invariant and uniquely characterise its shape. The same idea extends naturally to arbitrary configurations of points in Euclidean space. More generally, every physical measurement is ultimately a comparison between quantities of the same dimension. A length, for example, is determined only by comparing it with another length taken as a standard; similarly, a time interval is measured by comparison with the period of a clock, and a mass by comparison with a reference mass.\footnote{An article concerning dimensions, measurement and physical constants is currently in preparation. \textcolor{black}{Units such as the metre are arbitrary human conventions and cannot be applied meaningfully throughout the history of the universe. We will show how measurement can be formulated without any such arbitrariness.}} Consequently, the outcomes of measurements are dimensionless ratios rather than absolute quantities. For a closed system such as we assume the universe to be, however, there is no external standard against which an overall size could be defined. Therefore, we argue that only scale-invariant ratios are physically meaningful. In cosmology, for instance, the expansion of the universe is inferred from the changing ratio between intergalactic separations and characteristic galactic scales, while the cosmological redshift is itself defined as the ratio between observed and emitted wavelengths. Once one begins to look for them, such ratios appear to be ubiquitous.

Why then has physics not taken ratios, and therefore scale invariance, as one of its foundational principles? We argue that three conceptual structures, introduced formally by Newton, have profoundly shaped and, in a sense, constrained, our understanding of the world: absolute space together with absolute time, and an implicit notion of absolute scale, all rooted in the same reductionist mode of thought. Our motivation, therefore, is to free our theories from the absolute structures introduced by Newton and to develop a holistic theory in which scale invariance is elevated to the status of a first principle. We start with a $N$-body universe.

The principle of sufficient reason (PSR), formulated by Leibniz, provides a natural criterion for eliminating some of the absolutes. It states that every physical fact should admit a sufficient reason. In general, identifying such reasons is tremendously difficult. In the present context, however, the special scale-invariant values 0 and $\infty$ play a crucial role. Consider first the total energy. Choosing $E=\infty$ leads to a universe that disperses instantaneously, while any finite non-zero value is arbitrary because then energy carries dimensions and is therefore not scale-invariant. By contrast, $E=0$ is itself scale-invariant and yields perfectly well-defined solutions. The PSR therefore singles out universes with vanishing total energy. Exactly the same argument applies to the total angular momentum. Since any non-zero value is arbitrary, the PSR selects $\mathbf{L}=0$. This corresponds to a universe possessing no preferred rotational state. In addition, owing to Galilean invariance, one can always set the total linear momentum to zero, $\mathbf{P}=0$. The ontology is thereby reduced to a minimum, yielding a corresponding gain in conceptual clarity. The only absolute that remains is scale, and the remainder of this article is devoted to eliminating it.

\section{Variety} 

Consider a distribution of $N$ particles\footnote{We plan to extend the framework to an infinite number of particles.} with masses expressed as fractions of a total mass, $M$, normalised to unity, and Cartesian coordinates, $\mathbf{r}_i$, in space. We want to find a scale-invariant function that, taking into account all particles on an equal footing, can be used to characterise the extent to which they are uniformly distributed or clustered. Taking inspiration from Leibniz's \textit{Monadology}, where perfection is associated with the greatest possible variety compatible with the greatest order, we shall call this quantity the \emph{variety}.\footnote{Also referred to as the \emph{complexity}; see, for example, \cite{JB1, JB2}.}

Given our minimal ontology of point particles and separations, one of the simplest ways to construct a scale-invariant quantity is to take the ratio of two lengths. We choose as candidates the root-mean-square length, $\ell_\mathrm{rms}$, defined by
\begin{equation}\label{eq1}
	\ell_{\mathrm{rms}} = \sqrt{\sum_{i<j}m_im_jr^2_{ij}},
\end{equation}
and the mean-harmonic length, $\ell_\mathrm{mhl}$, given by
\begin{equation}\label{eq2}
	\ell^{-1}_\mathrm{mhl} = \sum_{i<j} \frac{m_im_j}{r_{ij}},
\end{equation}
where $r_{ij}=||\mathbf{r}_i-\mathbf{r}_j||$ denotes the distance between particles $i$ and $j$. 

The variety is then introduced as the dimensionless ratio
\begin{equation}\label{eq3}
	V = \frac{\ell_\mathrm{rms}}{\ell_\mathrm{mhl}}.
\end{equation}
The behaviour of the two characteristic lengths is markedly different under clustering: bringing a small subset of particles close together substantially decreases $\ell_\mathrm{mhl}$, thereby increasing $V$, while having only a minor effect on $\ell_\mathrm{rms}$. As a result, the variety provides a sensitive measure of clustering.

Other choices of characteristic lengths are certainly possible. The root-mean-square and mean-harmonic lengths are adopted here because they admit simple and compelling physical interpretations: the former is related to the square root of the centre-of-mass moment of inertia, $\sqrt{I_\mathrm{cm}}$, which provides a natural measure of the overall size of the system, while the latter is proportional to the negative inverse of the Newtonian gravitational potential, $V_\mathrm{New}$, when the gravitational constant is set to $G=1$. Thus, the two lengths are directly associated with the quantities that play a central role in Newtonian gravitational dynamics.

Hence, the variety can be written as:
\begin{equation}\label{eq4}
	V = - \sqrt{I_\mathrm{cm}}\,V_\mathrm{New}.
\end{equation}
Within the $N$-body literature, $V$ is commonly referred to as the shape potential or the normalised Newtonian potential. Whereas $V_\mathrm{New}$ is homogeneous of degree $-1$, the variety is the product of homogeneous functions of degrees $+1$ and $-1$, and is therefore homogeneous of degree $0$. Unlike the Newtonian potential, which depends on the overall scale of the configuration, the variety depends solely on its shape.

\section{Central Configurations}

A configuration of $N$ gravitating bodies is said to be a \emph{central configuration} (CC) if the gravitational force acting on each particle $i$ satisfies
\begin{equation}\label{eq5}
\sum_{j\neq i} Gm_im_j\,\frac{\mathbf r_i-\mathbf r_j}{r_{ij}^{3}}
=
\lambda m_i\left(\mathbf r_i-\mathbf R_{\mathrm{cm}}\right),
\end{equation}
where $\mathbf R_{\mathrm{cm}}$ is the centre of mass and $\lambda$ is a constant.

Central configurations have been extensively studied in the $N$-body problem. They generate homothetic and homographic solutions, in which the shape remains fixed while the system expands, contracts, or rotates, and they are precisely the critical points of the Newtonian potential restricted to configurations of fixed moment of inertia. Equivalently, and more relevant for our purposes, they are the critical points of the variety, $V$. Since the variety is the scale-invariant product of two homogeneous functions, each factor defines a force through its gradient whose strength is determined by the instantaneous value of the other. At every critical point these forces are exactly balanced, irrespective of the overall scale of the configuration. This is why central configurations are also known as \emph{relative equilibria}.

Whether the number of central configurations is finite or infinite for a given finite $N$ is the content of Stephen Smale's sixth problem for the twenty-first century \cite{S}. With chiral configurations distinguished, there are four central configurations in the three-body problem: 1 equilateral triangle and three collinear configurations. The first significant progress for $N=4$ and $N=5$ was made by Albouy and Kaloshin \cite{A}. The variety therefore possesses a vast number of critical points (more than factorial in $N$, and possibly infinitely many) consisting of minima and saddles. 

\begin{figure}[ht]
    \centering

    \begin{subfigure}{0.48\textwidth}
        \centering
        \includegraphics[width=\linewidth]{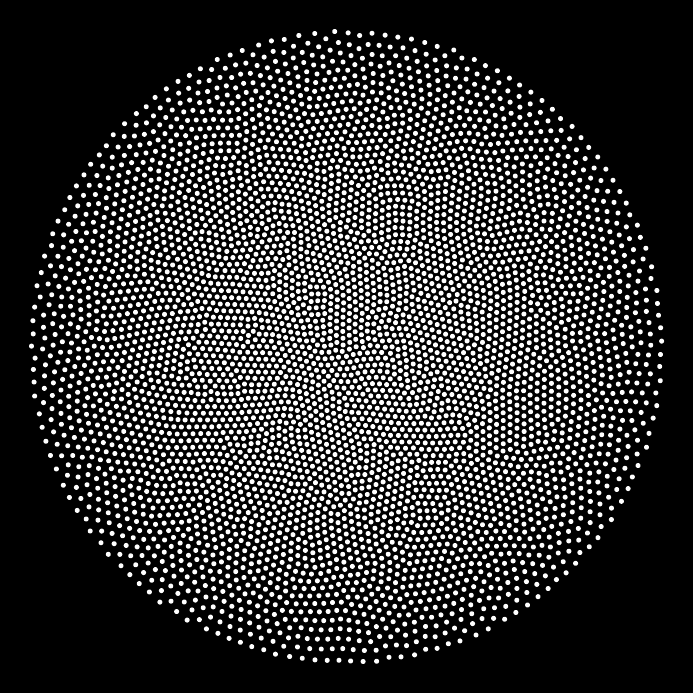}
    \end{subfigure}
    \hfill
    \begin{subfigure}{0.48\textwidth}
        \centering
        \includegraphics[width=\linewidth]{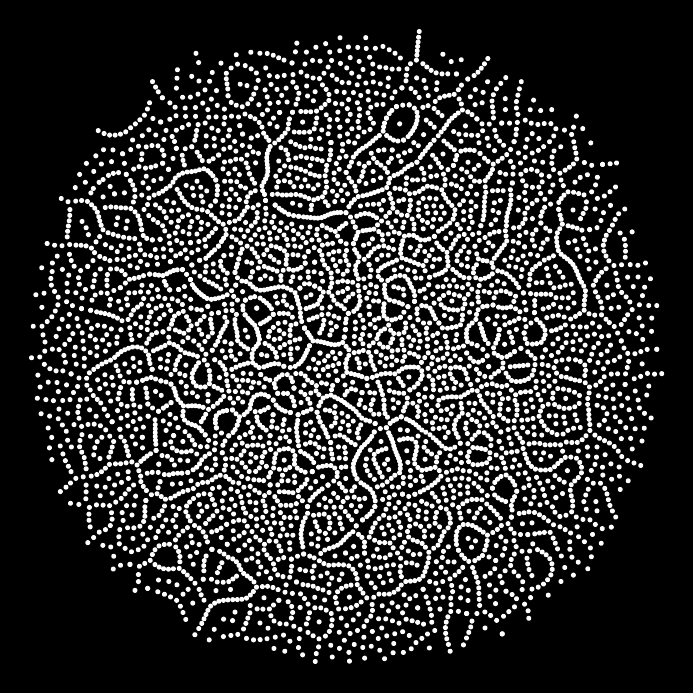}
    \end{subfigure}
    \caption{Two planar central configurations of 5,000 equal-mass particles. The configuration on the left lies at, or very close to, the absolute minimum of the variety, with V=0.585701. The configuration on the right corresponds to a higher critical point, with V=0.595574.}
    \label{fig1}
\end{figure}

Central configurations can be found numerically with relative ease. A well-established result in the $N$-body theory is that the variety possesses an absolute minimum  corresponding to exceptionally uniform particle distributions \cite{BG}. Until recently, numerical studies were primarily concerned with locating this global minimum, or configurations lying as close to it as possible, while avoiding local minima and saddle points. Our programme of shapes has motivated a recent shift in attention towards the wider structure of the variety and the critical points beyond its absolute minimum. In 2021, Manuel Izquierdo discovered that central configurations exhibit a striking filamentary structure away from the global minimum. He demonstrated this numerically in two-dimensional systems of 1,000 particles \cite{I}. Subsequent numerical studies have extended his results to larger systems in two dimensions and to three-dimensional configurations (see, for example, \cite{M1, M2}), revealing, in the latter case, filamentary structures of varying lengths, although no closed loops have been found. For our purposes, however, Fig.~\ref{fig1}, which shows two such critical configurations for a system of 5,000 particles, is sufficient to illustrate the essential qualitative behaviour. The configuration on the left lies very close to the absolute minimum of the variety, whereas the one on the right has a value of $V$ approximately $1.7\%$ above it. 

The filamentary structures shown in Fig.~\ref{fig1} emerge when $V$ is only slightly greater than its absolute minimum. In addition to these filaments, one observes closed loops containing several particles, a wealth of small-scale substructures, and particles distributed approximately uniformly throughout the filamentary network. Figure~\ref{fig1} bears an intriguing qualitative resemblance of the large-scale structure and evolution of our Universe. The configuration on the left is remarkably uniform, whereas the one on the right exhibits a filamentary network that bears a striking qualitative resemblance to the observed cosmic web. The increase in the separations between neighbouring particles from the centre towards the periphery of the configurations is an artefact of working in two dimensions, where Newton's potential theorem does not apply. Much remains to be explored. The rich landscape of the variety offers numerous directions for future work, including the study of its critical points in two and three dimensions for larger numbers of particles with the development of quantitative and qualitative measures of the structures that emerge, and a systematic comparison with established physical principles, such as the cosmological principle.

But the significance of central configurations does not end here. They play a central role in what are known as total-collision solutions: special Newtonian solutions in which the external scale $\sqrt{I_\mathrm{cm}}$ vanishes at a certain stage of the evolution, beyond which no unique continuation of the solutions has been found and Newton's equations are assumed to break down. For a total collision to occur, two conditions must be satisfied. First, the total angular momentum must vanish (there is no restriction on the energy). Second, the system must approach a very particular shape: a central configuration. In the case of three particles, in one scenario, the particles remain collinear, with the three possible orderings. In the more remarkable scenario, the triangle formed by the particles approaches an equilateral shape. Moreover, by time reversal of Newton's equations, total-collision solutions can be seen as Newtonian Big Bangs.

The requirement that the terminal shape be a critical point reveals that the term \emph{total collision} is somewhat misleading, as it reflects the intuition of an absolute scale inherited from Newton's absolute space and time. Although $\ell_\mathrm{rms}$ vanishes, this does not correspond to an event that internal observers could register, but rather to the approach towards a special shape---a central configuration. To make this clearer, we note that the variety is not only a measure of structure but also of the intrinsic scale of a particle distribution.\footnote{\textcolor{black}{Measurement of distance is now independent of a human choice of unit; it is entirely intrinsic. The universe itself provides the reference scale: every separation may be regarded as a ratio to the largest characteristic separation in the complete configuration, rather than to an externally chosen unit such as the metre.}} Length is determined operationally by comparing one distance with another: in the ideal limit of perfect accuracy, a length $R$ is measured by counting how many units of length $r$ fit into it as $r/R\rightarrow0$. In a many-body configuration, $\ell_\mathrm{rms}$ characterises the typical large separations between particles, whereas $\ell_\mathrm{mhl}$ is dominated by the smallest ones. Their ratio, the variety, therefore measures the hierarchy between large and small scales within the configuration. Although the extrinsic scale has been eliminated, the intrinsic scale encoded in the relative distribution of distances remains. 

It is also important to mention the counterparts of total-collision solutions, which are known as parabolic-escape solutions. In these solutions, the total energy must vanish, while the angular momentum may take any value. They also terminate in a configuration for which the variety has a critical value. However, in the $N$-body representation, rather than all particles colliding with infinite velocity at a single point and with zero separation between them, the particles in parabolic-escape solutions come to rest infinitely far apart.

These two classes of solution are already highly special. However, if only shape has physical significance, how should one interpret total-collision solutions with zero energy or parabolic-escape solutions with zero angular momentum? We argue that the distinction between a total collision and a parabolic escape exists only in the Newtonian description, where the scale variable $\sqrt{I_\mathrm{cm}}$ is still regarded as physically meaningful. Once scale is removed from the ontology, the distinction between the two solutions disappears.\footnote{\textcolor{black}{At the level of shape, both classes of solution terminate at a critical configuration of the variety. Although the succession of shapes completely specifies the objective content of the solution, its Newtonian representation requires additional arbitrary choices, such as the choice of initial coordinates and the origin and rate of time.}} What remains is simply a succession of shapes terminating at a critical point of the variety. From this succession of shapes alone, one could not determine whether it represents a total collision or a parabolic escape. A central configuration specifies only the shape of the system; it carries no information about the overall scale at which that shape is realised. Consequently, the same terminal shape may equally well be interpreted as a Newtonian Big Bang, for which $\sqrt{I_\mathrm{cm}}=0$, or as the end of a parabolic escape, for which $\sqrt{I_\mathrm{cm}}\rightarrow\infty$. Zero and infinity are opposite limits of the same quantity. Once that quantity is removed from the ontology, they can no longer distinguish physically different histories. From this perspective, Newton's equations exhibit not only time-reversal symmetry but also a symmetry under inversion of scale---a symmetry that may have remained concealed by the absolutes Newton introduced in his theory.

But there is more that is hidden by the absolutes. As is generally acknowledged, the entropic arrow of time faces several conceptual difficulties. One of them is that it requires the universe to have had, at some time in the past, an extraordinarily low entropy for which the known laws of nature provide no explanation. In \cite{JB3} it is shown that all solutions for which the energy is non-negative have non-entropic bidirectional arrows of time aligned with the secular growth of the variety either side of a \emph{Janus point};\footnote{The Janus point is the unique minimum of $I_{\mathrm{cm}}$ along a solution, from which bidirectional arrows of time arise.} they are significantly more pronounced if the energy and angular momentum are both zero \cite{JB1} and are expressed even more clearly in the solutions that could be either of the total-collision or parabolic-escape type. In other words, we find a direction of time defined by increasing structure and variety in all the solutions allowed by our principles. On that basis, we can say which of any two shapes comes first in time as revealed by actual structure, rather than by Newton's absolute time, which possesses no intrinsic physical attribute. In the scale-invariant shape-based approach shapes of ever changing structure, not entropy, define direction. Shapes are the true instants of time.

We have already noted that it is not known whether the number of relative equilibria for a given $N$ is finite or infinite. At the very least, there are a great many---already factorial in $N$ for the collinear configurations alone. Since the direction of time is defined by increasing variety, every relative equilibrium may be regarded as a possible first instant of a solution that, in Newtonian terms, corresponds either to a total collision or to a parabolic escape.
Can this still vast, and possibly infinite, family of candidate solutions be reduced? Here the PSR can again be invoked. If there is to be no arbitrary choice, only those solutions whose first instant has the absolute minimum value of the variety should be admitted. However, even this criterion may fail to select a unique solution. Except for relatively small values of $N$, it is not known whether the absolute minimum of the variety is unique. If several distinct shapes realise the same minimum, they are expected to possess statistically equivalent uniformity properties. In that case, the PSR can take us no further: it reduces the arbitrariness as far as possible, but may not eliminate it entirely.

Besides the qualitative ordering of shapes, we also suggest a quantitative ordering by defining the age $a(s)$ of any shape $s$ of $N$ particles with given mass ratios as the excess of its variety over that of the shape (or shapes) with the absolute minimum $V_0$ for the same particles:
\begin{equation}\label{eq6}
a(s)=V(s)-V_0.
\end{equation}
In Fig.~\ref{fig1}, the configuration on the left lies at, or extremely close to, the absolute minimum of the variety, whereas the one on the right has a value approximately $1.7\%$ greater. Their corresponding ages are therefore $0$ and $0.017$.\footnote{\textcolor{black}{The notion of age is not restricted to the universe as a whole. Any sufficiently well-defined substructure may likewise be assigned an age according to the amount of structure it contains. In two-dimensional configurations, however, the identification of such substructures is often ambiguous, whereas in the three-dimensional central configurations found so far the filamentary structures appear to be uniquely distinguished by their lengths \cite{M2}.}} Unlike conventional notions of age, such as ``13.8 billion years since the Big Bang'', which are expressed in units ultimately defined by the Earth's orbital motion---even though the Earth did not always exist---this quantity measures the amount of structure relative to the most uniform possible configuration. 
\textcolor{black}{Age may therefore be regarded as an alternative to duration, which Newton used as another name for absolute time.}

Physics has the tradition of explaining phenomena by referring to a succession of events in time. We shall call this \emph{explanation through time}. Here we explore an alternative possibility: that explanation may instead be encoded in the structure of a single shape. We call this \emph{explanation through space}.

Fossils, for example, are present-day records of organisms that lived in earlier ages of the Earth. The experimental evidence from particle accelerators such as CERN reaches physicists only in the form of detector images or computer outputs. More generally, all evidence about the past and about the laws of nature is available only through present physical records, which are themselves nothing but particular configurations of particles. Likewise, the resemblance between the particles outside the filamentary network (Fig.~\ref{fig1}, right) and the uniform configuration of minimum variety (Fig.~\ref{fig1}, left) suggests that the former preserves information about the latter, which may be interpreted as \emph{records} of the \emph{first instant}. As variety increases, structure accumulates, so that each shape contains traces of less complex configurations. From this perspective, the past is not something that exists \emph{behind} the present in an external time dimension, but is encoded in the internal structure of a single configuration that we call the present. That is, all the past may be encoded as complexity in the present. The present thus becomes a repository of records from which both the past and the laws governing the universe can be inferred.

\section{Final Discussion and Conclusion}

Before concluding, we briefly comment on general relativity and quantum mechanics, beginning with the former. Conceptually, apart from the role ascribed to light,\footnote{We are currently preparing a separate article in which the role of light is discussed in detail.} the ontology of physics has changed less than is often supposed. Minkowski replaced Newton's separate notions of absolute space and time with a unified spacetime, but the underlying ontological picture remained essentially unchanged. Just like Newton, he relied on an invisible structure---in his case, spacetime---to explain observable phenomena. General relativity represented a profound advance by replacing the fixed geometry of Minkowski spacetime with a dynamical one. Nevertheless, the theory still presupposes the existence of a differentiable spacetime manifold that is locally Minkowskian. It is the \emph{glue} that, everywhere and everywhen, holds spacetime together. Yet, apart from the causal structure defined by light cones, it is featureless. Moreover Einstein himself admitted \cite{E1} to have committed a \emph{sin} in his creation of general relativity as an incomplete theory that postulates the existence of two critical entities: the spacetime metric, and rods and clocks brought in to measure the metric rather than emerging from the fundamental equations
of the theory.

But there is more. Poincaré argued that Euclidean geometry could, in principle, always be retained, first for the sake of simplicity, and second provided that apparent deviations from it were interpreted as the effects of physical interactions on measuring instruments rather than as intrinsic curvature of space itself.\footnote{Einstein himself later recognised in \cite{E2} that such reformulations were logically possible, even though he regarded the geometric formulation of general relativity as more natural.} The geometry we attribute to physical space is therefore not an \textit{a priori} necessity, but an effective description grounded in empirical measurement. Rigid rods and clocks, being physical systems, are themselves subject to forces such as gravitation and electromagnetism, so their properties---and the measurements they yield---depend on the local physical conditions. General relativity deepened Poincaré's insight regarding the operational nature of geometry. In the presence of gravitational fields, rods contract and clocks dilate, so that the geometry inferred from measurements depends directly on the distribution of mass and energy. The geometry attributed to the world therefore cannot be separated from the physical behaviour of the systems used to measure it.

This leads to an important distinction between two notions of geometry. On the one hand, there is the \emph{background geometry}---the mathematical structure assumed by a theory, such as Euclidean space in Newtonian mechanics. On the other, there is the \emph{measured geometry}---the geometry inferred from physical systems embedded in that background. The two need not coincide. Distances, for example, are apprehended through physical procedures involving rods, clocks, light signals, or the relative separations of particles. Every such method is constrained by the interactions acting upon the measuring apparatus itself. In self-gravitating systems, gravity naturally gives rise to spatial inhomogeneities in the distribution of matter, and these in turn shape the relative separations from which geometric properties are inferred \cite{M2}. What emerges, therefore, even within Newtonian gravity, is an effective geometry that may vary from place to place---not as an anomaly, but as a natural consequence of defining geometry through physical measurement.

A further motivation for reconsidering the ontology of general relativity comes from the remarkable richness of its solution space. The Einstein field equations admit an enormous variety of mathematically consistent solutions, many of uncertain physical significance. By contrast, the relational analysis of the $N$-body problem suggests that many typical Newtonian solutions can be eliminated through relational and scale-invariant principles. This raises the possibility that a comparable \emph{purification} of general relativity should be possible. If such a programme can be realised, many mathematically admissible but physically questionable solutions of Einstein's equations may cease to arise, with potentially far-reaching consequences for both gravitation and quantum theory.

Although our article has been about gravity, the problem with reductionism and the boundary and initial conditions it enforces unnaturally on the universe can just as well apply to quantum mechanics. Indeed, many physically relevant solutions cannot be obtained without boundary conditions. For example, the possible states of the hydrogen atom can only be obtained by requiring its wave function to vanish at spatial infinity. From our perspective, boundary conditions are symptoms of an incomplete ontology: they compensate for describing the universe through a reductionist lens. By contrast, our relational programme treats the universe as a holistic whole. The admissible solutions are selected by intrinsic relational principles rather than by externally imposed boundary conditions. Whether the same principle can be extended to quantum mechanics remains an open question, but the present work suggests that such an extension may well be possible.

One reason for this optimism is that timeless correlations in quantum mechanics are often regarded as mysterious, yet analogous correlations arise already in Euclidean geometry, although they are of a different nature. A configuration of $N$ points possesses $N(N-1)/2$ pairwise separations, but only $3N-7$ independent degrees of freedom in three-dimensional space, if we remove translations, rotations, and dilatations. The remaining pairwise separations are therefore constrained by purely geometric consistency relations. Physicists rarely regard these correlations as mysterious because they are accustomed to working within Euclidean geometry. From our perspective, however, they illustrate a more general principle: the structure of a configuration can itself encode non-trivial correlations without any appeal to temporal evolution. The correlations encountered in quantum mechanics may therefore be less exceptional than is commonly supposed. A final conceptual remark concerns time. Despite its radical departure from classical mechanics, it is remarkable that, in standard quantum mechanics, time is treated exactly as Newton conceived it: as an external, featureless, absolute parameter.

There is no doubt that Newtonian mechanics, general relativity, and quantum mechanics rank among the greatest achievements of science and have been confirmed to extraordinary precision. Nothing in the present work is intended to diminish their empirical success. Yet it remains possible that the principles on which they are founded do not provide the most faithful description of nature. The programme presented here is still at an early stage, but it already suggests that a minimal ontology, combined with relational, holistic, and scale-invariant principles, uncovers structures obscured by the conventional formulation. Removing the absolute structures introduced by Newton appears to endow Newtonian gravity itself with an unexpected richness: rather than becoming poorer, the theory acquires a remarkable creative\footnote{The word \emph{creative} is used here in its etymological sense of \emph{growth}.} potential. This suggests that those absolute structures are ultimately not needed to explain the universe.

We have long been accustomed to describing reality in terms of absolute structures. This habit is deeply rooted in our experience on Earth, where we can always appeal to an external background against which positions, motions, and scales are defined. When the universe as a whole is considered, however, no such external backdrop exists. If the same principles can be successfully extended to general relativity and quantum mechanics, we may ultimately gain far more by removing than by adding.

\section*{Acknowledgements}

\textcolor{black}{M.~I.~R.~L. gratefully acknowledges the Instituto de Astrofísica e Ciências do Espaço (IA) for providing access to the computing cluster used in this work. J.~B. thanks Tim Koslowski for many helpful discussions.}

\end{document}